\documentclass[letterpaper]{article}
\usepackage{aaai}
\usepackage{times}
\usepackage{helvet}
\usepackage{courier}
\usepackage{url}
\usepackage{graphicx}
\usepackage{amssymb}
\begin{document}
%
\title{Dropout Prediction in Crowdsourcing Markets}
\author{Malay Bhattacharyya\\
Department of Information Technology\\
Indian Institute of Engineering Science and Technology, Shibpur\\
Howrah -- 711103, India\\
E-mail: malaybhattacharyya@it.iiests.ac.in}
\maketitle
\begin{abstract}
\begin{quote}
Crowdsourcing environments have shown promise in solving diverse tasks in limited cost and time. This type of business model involves both the expert and non-expert workers. Interestingly, the success of such models depends on the volume of the total number of workers. But, the survival of the fittest controls the stability of these workers. Here, we show that the crowd workers who fail to win jobs successively loose interest and might dropout over time. Therefore, dropout prediction in such environments is a promising task. In this paper, we establish that it is possible to predict the dropouts in a crowdsourcing market from the success rate based on the arrival pattern of workers.
\end{quote}
\end{abstract}

\section{Introduction}
The recent emergence of crowdsourcing environments has shown immense success in solving diverse tasks in a limited time and bounded cost \cite{Brabham2013}. It has created a new kind of business model that involves both the expert and non-expert online workers (crowd) to get the job done. It is statistically realizable that the success of such models highly depends on the volume of the crowd. On the other hand, the stability of the crowd in this volume is dependent on their consistent performance. If the consistency decreases the dropout rate might increase. Therefore, identifying the potential dropouts in such environments has a major promise. No significant attempt has been made earlier in this direction. In the current paper, we primarily explore the dependence between the winning pattern and the rate of survival in a crowdsourcing platform. We also verify whether it is possible to predict the dropouts in a crowdsourcing market from the performance of the workers.

\section{Related Works}
The prediction of dropouts has been extensively studied in the context of academic association \cite{Oreopoulos2007}. This mainly encompasses the study of student dropouts from different perspectives. It has been recently extended to predict the dropouts in MOOC platforms too \cite{Halawa2014}. However, rare attempts have been made to study the cause of dropouts in a labor market by using learning mechanisms. The limited existing approaches focus on the effect of eduction \cite{OECD2011}, skills and work addiction \cite{Muller2010}, and personality characteristics \cite{Viinikainen2014} of workers on the rate of dropouts. However, there exists no approach for the prediction of dropouts in crowdsourcing markets. In this paper, we study whether a prediction model could be used for identifying the dropouts in such environments.

\section{Preliminaries}
We assume that the jobs and the crowd workers arrive into the crowdsourcing market following different probability distributions. The arrival time of a crowd worker is the time when the worker participates into a specific job. Given the set of arrival times of a particular worker, say $\mathcal{T} = \{t_1, t_2, \ldots, t_n\}$, the {\em inter-arrival} time can be defined as $\triangle \mathcal{T} = \mathcal{T}\diagdown t_1 - \mathcal{T}\diagdown t_n$. It basically reflects the set of time gaps between two successive participations in a given time frame. We demarcate a crowd worker as {\em dropout} if $\triangle \mathcal{T}_{n-1} > \psi$, i.e., the final inter-arrival time is strictly more than a threshold.

\section{Dataset Details}
Flightfox is a crowdsourcing environment that involves crowd workers to find out the best itineraries for the requesters \cite{Flightfox}. It posts every task (for searching the suitable itinerary) as a contest to be solved through crowdsourcing. We analyze this crowdsourced data, comprising 13,114 contests completed between a period of 20 months. Some necessary terminologies associated to this platform are contests (a task to be solved), flyers (the requesters), experts (the crowd workers), winners (the crowd worker who submits the best solution), and the finders fee' (the incentive). The collected data includes a set of 1067 crowd workers in total.

\section{Preliminary Results}
We primarily put together the entire data to construct two separate networks, namely a participation network (between the workers and the tasks) and a winner network (between the winners and the tasks). We compute the degree values of all workers in the participation and winner networks. Our goal is to find whether there exists any relation between the participation and winning pattern. We also calculate the success rate (= winning degree/participation degree) of each worker from these networks. We then compute the Pearson correlation coefficient ($\rho$) between the winning degree and the participation degree. It is found that the dependence is positive and is fairly strong ($\rho = 0.88$). Evidently, the more a worker participates, the more he has a chance to win.

With this understanding, we now try to recognize which workers might be dropping out from participating in future tasks. For this, we first need to find out the dropout workers from our given data. We divide the list of workers into two separate lists, as per their order of participation, with a ratio of 2:1. The first two-third of the workers form the training set and the remaining one-third form the test set. We consider the dropout workers as those who have participated in at least one job in the training set but have not participated in any job in the test set. There were 724 such dropout workers as identified by the said principle.

We first calculate the success rate of these dropout workers based on their participation in the training list. The whole list of dropouts is divided into a number of ranges, i.e., we tabulate the number of dropouts with success rates between 0-10\%, 11-20\%, ..., 91-100\%. Based on this, we calculate the mean value of success rates in each range. Thus, we obtain a pair of observations, namely the mean success rate of a range and the corresponding number of dropouts having that success rate. These results are shown in details in Table~\ref{Table1}.

\begin{table}
  \centering
  \begin{tabular}{|c|c|c|}
    \hline
    \textbf{Range of} & \textbf{Number of} & \textbf{Average success} \\
    \textbf{success rate} & \textbf{dropouts} & \textbf{rate (\%)} \\
    \hline
    0-10 & 366 & 0.15 \\
    \hline
    10-20 & 43 & 15.64 \\
    \hline
    20-30 & 62 & 25.40 \\
    \hline
    30-40 & 72 & 35.58 \\
    \hline
    40-50 & 78 & 48.97 \\
    \hline
    50-60 & 15 & 56.20 \\
    \hline
    60-70 & 14 & 66.33 \\
    \hline
    70-80 & 8 & 76.89 \\
    \hline
    80-90 & 2 & 84.60 \\
    \hline
    90-100 & 64 & 100 \\
    \hline
  \end{tabular}
  \caption{Number of dropouts versus the success rate.}
  \label{Table1}
\end{table}

As can be seen from Table~\ref{Table1}, the final range of success rate (90-100\%) highlights a sudden increase in the number of dropouts. This is possibly because these workers have participated in a very less number of tasks ($\leq 7$) and have won most of them. Therefore, the success rate is quite high. We have not considered them in calculating the coefficient. But even if they are included, the coefficient remains same. The correlation coefficient is roughly obtained as $-0.73$, which is a strong negative dependence. So, it appears that lesser the success rate, more is the dropout tendency. At this point, we apply a number of basic classification algorithms to be trained from the data to predict whether a worker is a probable dropout or not. The k-NN and Bayes classifier algorithms are used to train the data (two-third of the workers). The features that we considered are the participation degree, winner degree, and success rate. Finally, we test the model with the remaining one-third data.

Using both the k-NN and Bayes classifiers, the prediction of dropout is achieved with accuracies of close to 70\%. We also apply the cross-validation method to better evaluate the results. We first divide the whole data into different ratios of training and test sets, i.e., 10\% train set-90\% test set, 20\% train set-80\% test set, ..., 90\% train set-10\% test set. We then apply the k-NN and Bayes classifiers on each set. The detailed results are shown in Table~\ref{Table2}. We see that there is a significant effect of this separation ratio (of training and test sets) on the classifier performance. However, the classification accuracy is obtained as high for all the cases.

\begin{table}
  \centering
  \begin{tabular}{|c|c|c|c|}
    \hline
    \textbf{Train-Test} & \textbf{k-NN (k = 1)} & \textbf{k-NN (k = 3)} & \textbf{Bayes} \\
    \hline
    10-90 & 66.00 & 69.45 & 74.24 \\
    \hline
    20-80 & 66.35 & 69.17 & 74.33 \\
    \hline
    30-70 & 65.86 & 70.55 & 73.90 \\
    \hline
    40-60 & 67.03 & 71.25 & 74.22 \\
    \hline
    50-50 & 69.23 & 71.67 & 75.80 \\
    \hline
    60-40 & 67.91 & 70.96 & 75.88 \\
    \hline
    70-30 & 68.75 & 71.88 & 77.19 \\
    \hline
    80-20 & 69.48 & 71.83 & 77.00 \\
    \hline
    90-10 & 67.29 & 75.70 & 79.44 \\
    \hline
  \end{tabular}
  \caption{Performance in classifying the dropouts.}
  \label{Table2}
\end{table}

\section{Future Scope}
This paper highlights that success rate, defined over the inter-arrival time, is a promising feature in predicting the dropouts in crowdsourcing markets. Although having used simple classifiers, we obtain a high classification rate. So, it is possible to devise additional features and robust classification models for a better prediction of dropouts.

\section{Acknowledgment}
The author would like to thank Smritikona Barai of Heritage Institute of Technology for her preliminary support.

\bibliographystyle{aaai}
\bibliography{Reference}

\end{document}